\documentclass[
 reprint,amsmath,amssymb,aps,prl,
]{revtex4-1}

\usepackage{graphicx}
\usepackage{dcolumn}
\usepackage{bm}
\usepackage{textcomp} 

\begin{document}

\title{Squeezed vacuum states from a whispering gallery mode resonator}

\author{Alexander Otterpohl$^{1,2,3,*}$\email{Alexander.Otterpohl@mpl.mpg.de}}
\author{Florian Sedlmeir$^{1,2}$}
\author{Ulrich Vogl$^{1,2}$}
\author{Thomas Dirmeier$^{1,2}$}
\author{Golnoush Shafiee$^{1,2}$}
\author{Gerhard Schunk$^{1,2,3}$}
\author{Dmitry V. Strekalov$^{1}$}
\author{Harald G.L. Schwefel$^{4,5}$}
\author{Tobias Gehring$^{6}$}
\author{Ulrik L. Andersen$^{1,6}$}
\author{Gerd Leuchs$^{1,2}$}
\author{Christoph Marquardt$^{1,2}$}
\affiliation{
$^{1}$Max Planck Institute for the Science of Light, Staudtsta\ss e 2, 91058 Erlangen, Germany\\
$^{2}$Institute of Optics, Information and Photonics, Friedrich-Alexander University Erlangen-N\"urnberg, Staudtstra\ss e 7 B2, 91058 Erlangen, Germany\\
$^{3}$Erlangen Graduate School in Advanced Optical Technologies (SAOT), Friedrich-Alexander University Erlangen-N\"urnberg, Paul-Gordan-Stra\ss e 6, 91052 Erlangen, Germany\\
$^{4}$The Dodd-Walls Centre for Photonic and Quantum Technologies, 730 Cumberland Street, 9016 Dunedin, New Zealand\\
$^{5}$Department of Physics, University of Otago, 730 Cumberland Street, 9016 Dunedin, New Zealand\\
$^{6}$Center for Macroscopic Quantum States (bigQ), Department of Physics, Technical University of Denmark, Fysikvej, 2800 Kgs. Lyngby, Denmark
}

\begin{abstract}
Squeezed vacuum states enable optical measurements below the quantum limit and hence are a valuable resource for applications in quantum metrology and also quantum communication. However, most available sources require high pump powers in the milliwatt range and large setups that hinder real world applications. Furthermore, degenerate operation of such systems presents a challenge. Here, we use a compact crystalline whispering gallery mode resonator made of lithium niobate as a degenerate parametric oscillator. We demonstrate about 1.4 dB noise reduction below the shot-noise level for only 300 \textmu W of pump power in degenerate single-mode operation. Furthermore, we report a record pump threshold as low as 1.35 \textmu W. Our results show that the whispering gallery-based approach presents a promising platform for a compact and efficient source for nonclassical light.
\end{abstract}

\maketitle

\section{Introduction}
Since the first experimental realizations of squeezed light in the 1980s \cite{PhysRevLett.55.2409}, there has been much research on different quantum systems in order to maximize the amount of available squeezing \cite{1402-4896-91-5-053001}. So far, the cavity-assisted generation of squeezed light based on parametric down conversion (PDC) called an optical parametric oscillator (OPO) \cite{PhysRevLett.57.2520} was shown to be the most efficient. The approach with monolithic PPKTP cavities delivered steady improvements in terms of squeezing \cite{PhysRevLett.104.251102}, stabilization \cite{Yonezawa:10}, and pump power \cite{Schonbeck:18}, culminating in the highest squeezing levels up to 15 dB \cite{PhysRevLett.117.110801}. The wide field of applications for squeezed light ranges from quantum communication schemes \cite{ncomm:tgehring} and quantum imaging \cite{Treps940} to quantum-enhanced metrology. The most prominent example for the latter is interferometric applications \cite{SCHNABEL20171} as used in gravitational wave detectors like GEO 600 \cite{nphys:ligo2011}. More recently, the continuous variable (CV) quantum computing application \cite{RevModPhys.77.513} raised a demand for a compact and low-power source for squeezed states of light in order to facilitate the needed scalability.\\
First on-chip results were achieved with $\chi^{(3)}$-nonlinear silicon nitride microring resonators that were used for generating two-mode squeezing above the threshold \cite{PhysRevApplied.3.044005,Dutt:optlett:41} and recently for two-mode vacuum squeezing below the OPO threshold \cite{arxiv:1904.07833}. But driving the underlying process of four-wave mixing degenerately \cite{Hoff:optexpress:23} in order to get genuinely single-mode squeezed vacuum states of light in degenerate operation has remained a challenge.\\
In this work, we meet the challenge of generating genuinely single-mode squeezed vacuum states of light in a crystalline whispering gallery mode resonator (WGMR), which generally is a more efficient source due to the $\chi^{(2)}$-nonlinear interaction. WGMRs provide an efficient and compact platform for nonlinear and quantum optics \cite{2040-8986-18-12-123002}. In particular, we have shown that WGMRs can be used for the generation of quantum-correlated signal and idler beams \cite{PhysRevLett.106.113901}, which exhibit individual intensity squeezing, and tunable heralded single photons \cite{ncomm:mfoertsch}. The WGMRs' high Q-factors yield very low parametric oscillation thresholds of a few microwatts \cite{PhysRevLett.105.263904} and provide high tunability over the whole transparency window from the UV up to the telecom regime \cite{doi:10.1080/09500340.2016.1148211}. However, operation in strictly degenerate mode is challenging, as phase-matching conditions put stringent requirements on the stability of the setup and thermorefractive noise \cite{Gorodetsky:04,Phys332} might impede vacuum squeezing.\\
Here, we show the operation of a $\chi^{(2)}$-nonlinear WGMR as a degenerate doubly resonant OPO below threshold. Our experiment showed about 1.4 dB of vacuum squeezing at a sideband frequency of 500 kHz requiring only 300 \textmu W of incident pump power at 532 nm. The current system provides two adjustable coupling ports for in- and outcoupling, which allows for active tuning of the parametric oscillation threshold and for using an additional alignment beam for the detection setup. The experimental approach consists of two steps. We first determine the parametric oscillation threshold of the OPO, and then we pump the system above and below the threshold. In general, the OPO has to be operated below the threshold for generating squeezed vacuum states of light. Our tunable low threshold enables us to investigate the above threshold regime as well.

\section{Theoretical model}
The system is theoretically modeled as a parametric oscillator within the input output formalism with two coupling ports that have independent coupling rates $\gamma_1$ and $\gamma_2$ (see Fig. \ref{fig:experimental_setup}) at the subharmonic frequency, as discussed in \cite{PhysRevA.30.1386}. The pump is treated classically without depletion, which is appropriate below the OPO threshold. The system is pumped through the first port. The noise variance of the squeezed (anti-squeezed) quadrature $X_\text{sq/asq}$ at sideband frequency $f$ is measured after the second port with homodyne detection and can be calculated as \cite{Polzik1992}
\begin{equation}
\text{Var}(X_\text{sq/asq})[f]= 1\mp 4\eta \mathcal{V}^2  \frac{\gamma_\text{2}}{\Gamma} \cdot \frac{\sqrt{\mathcal{G}}}{\left ( 1\pm \sqrt{\mathcal{G}} \right )^2 +  \left (\frac{2\pi f}{\Gamma}\right)^2}\, ,
\label{eq:squeezingquadraturenoisevariance}
\end{equation}
with $\mathcal{G}$ being the cooperativity, $\eta$ being the detection efficiency, $\mathcal{V}^2$ being the homodyne efficiency describing the spatial mode overlap between signal and local oscillator, and 
\begin{equation}
\Gamma = \frac{\Gamma_\text{FWHM}}{2} = \gamma_1+\gamma_2+\gamma_\text{int}
\end{equation}
being the total loss rate. The intrinsic loss of the resonator is described by the rate $\gamma_\text{int}$. The cooperativity 
\begin{equation}
\mathcal{G} = \frac{4g^2 n_\text{p}}{\Gamma^2}\, ,
\end{equation}
with $g$ being the nonlinear coupling constant and $n_\text{p}$ being the number of pump photons, is the ratio of nonlinear conversion and total loss.\\
A unity cooperativity $\mathcal{G} = 1$ marks the OPO threshold where the nonlinear conversion matches the loss. Being pumped through the first port, the threshold power of the system is given by
\begin{equation}
P_\text{th} = \frac{\hbar \omega_\text{p}}{8g^2}\, \frac{\Gamma^2\Gamma_\text{p}^2}{\gamma_{1\text{p}}}\, .
\label{eq:pumpthreshold}
\end{equation}
The index p indicates pump related parameters.\\
Above the threshold, a bright signal field is generated according to \cite{doi:10.1080/09500340.2016.1148211}
\begin{equation}
P_\text{signal} = 4\, P_\text{th} \frac{\gamma_{1}}{\Gamma} \frac{\gamma_{1\text{p}}}{\Gamma_\text{p}} \cdot \left ( \sqrt{\frac{P_\text{p}}{P_\text{th}}}  - 1 \right )\, ,
\label{eq:signalpower}
\end{equation}
for zero pump detuning from resonance and zero phase-matching detuning. This functional relation is later used to determine the OPO threshold in the experiment.\\
As a function of cooperativity, the squeezed quadrature variance reaches its minimum when 
\begin{equation}
\mathcal{G}=1+(2\pi f/\Gamma)^2>1\, . 
\end{equation}
This means that maximum squeezing is possibly reached above the threshold, given the assumption of a pump without depletion.
Usually, the sideband frequency is chosen to be much lower than the resonator linewidth, $f\ll\Gamma$, in which case the best squeezing is achieved just above $\mathcal{G}=1$. Then, for unity detector efficiency and perfect overlap with the local oscillator, $\eta \mathcal{V}^2 = 1$, the squeezed quadrature becomes $\text{Var}(X_\text{sq})=1-\gamma_2/\Gamma$. If the coupling rates are equal, $\gamma_1=\gamma_2$, the attainable squeezing level is limited to 3 dB. To attain a stronger squeezing, the first coupler should be undercoupled and the second strongly overcoupled. Then $\gamma_2\gg \gamma_1+\gamma_\text{int}$, and the squeezed variance becomes $(\gamma_1+\gamma_\text{int})/\gamma_2$.

\section{Experimental setup}
\begin{figure*}
\includegraphics{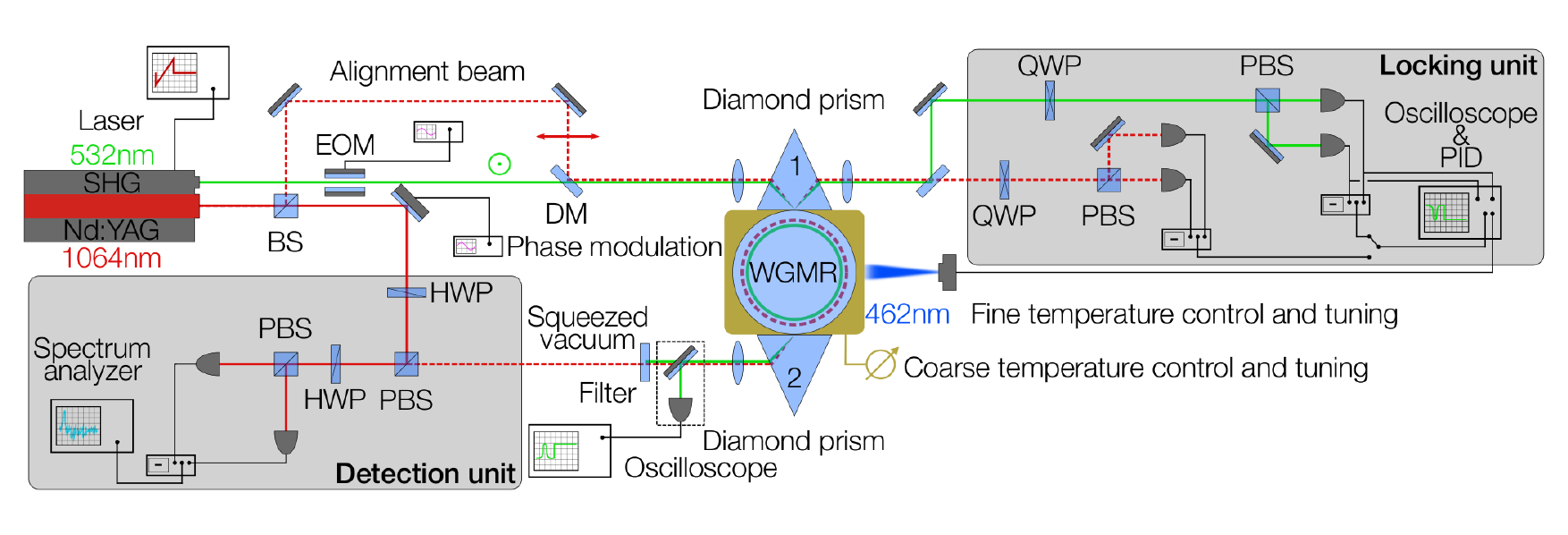}
\caption{Experimental setup. Squeezed vacuum states of light are generated via degenerate parametric down conversion of a 532 nm pump inside a crystalline whispering gallery mode resonator (WGMR). The prism couplers 1 and 2 have independent coupling rates $\gamma_1$ and $\gamma_2$ at the subharmonic frequency, respectively. The lock is based on the H\"ansch-Couillaud scheme and applicable to both the pump beam and the alignment beam. The homodyne detection is implemented with two polarizing beam splitters. The dashed box indicates a flip mirror for optionally detecting the intracavity pump power. BS, beam splitter; DM, dichroic mirror; EOM, electro-optic modulator; HWP, half-wave plate; PBS, polarizing beam splitter; PID, proportional-integral-derivative controller; QWP, quarter-wave plate; SHG, second harmonic generation.}
\label{fig:experimental_setup}
\end{figure*}
We use a WGMR with a major radius $R \approx 2.48\text{ mm}$ and a minor radius $r \approx 0.49\text{ mm}$ made of 5 mol.\,\% MgO-doped z-cut lithium niobate ($\text{LiNbO}_3$). Our resonator is fabricated with a single point diamond turning machine and afterwards polished in order to provide the required surface grade. We pump the system with a 532 nm Nd:YAG laser (Innolight, Prometheus) and deploy its fundamental wavelength at 1064 nm as an alignment beam, which is used for aligning the vacuum mode to the local oscillator and for exact tuning to the degenerate point of operation. Type-I phase-matching, which is thoroughly explained for such resonators in \cite{doi:10.1080/09500340.2016.1148211}, requires an ordinarily polarized alignment beam and an extraordinarily polarized pump beam. This setting results in a phase-matching temperature around $109^\circ\text{C}$ for degenerate PDC, which is coarse-controlled with a Peltier heater. A fine control of the resonator temperature is achieved by optically heating the resonator, as explained below.\\
As shown in Figure \ref{fig:experimental_setup}, we use a configuration with two coated diamond prism couplers for coupling evanescently to the resonator. Both prisms are mounted on piezo actuators with nanometer precision, which allows for continuous and independent adjustment of the coupling rates $\gamma_1$ and $\gamma_2$. The pump and alignment beams are coupled to the resonator through the first port. The second coupler allows us to directly monitor the intracavity pump power. It is also used to couple out the generated infrared light from the resonator without having the uncoupled residuals of the input beams. We can hereby check the spatial mode profile to ensure the fundamental mode operation \cite{Schunk:14}, which yields the highest overlap between the pump and parametric WGMs \cite{PhysRevLett.104.153901} and, consequently, the best nonlinear interaction strength. Besides that, the alignment beam greatly simplifies alignment of the local oscillator, allowing us to measure and optimize the homodyne efficiency $\mathcal{V}^2$. For this measurement, we balance the local oscillator power and the alignment beam power emitted from the second coupler, modulate the local oscillator phase, and measure the visibility $\mathcal{V}$ of the observed interference fringes. Using this technique, we measured $\mathcal{V}$ up to 92\%.\\
However, the parametric light is coupled more strongly by the first coupler than the pump beam because the evanescent field decays exponentially with the wavelength outside the resonator. This introduces an inevitable loss to the squeezed vacuum state and sets a limit for minimizing $\gamma_1$ because the pump beam needs to be coupled through the same port, which could be countered by using selective coupling \cite{PhysRevApplied.7.024029}.\\ 
We used the beams reflected from the first prism coupler to monitor the coupling of the pump and alignment beams to their respective WGMs and to implement a H\"ansch-Couillaud locking scheme \cite{HANSCH1980441} for each beam. We critically coupled the pump beam and measured the coupling efficiency, which is mainly influenced by the spatial mode overlap of the incident pump beam and the whispering gallery mode emitted from the resonator through the coupler. For an optimized beam and resonator, the coupling efficiency can exceed 99\% \cite{Strekalov:09}. In this experiment, we measured 75\%. In the following, we will multiply the incident pump power by this coefficient, referring to it as the incoupled pump power.\\
The locking scheme allowed us to lock the resonator either to the infrared mode for alignment, or to the pump mode for generating squeezed vacuum states of light. The error signal was processed and fed to a 462 nm laser diode, which is mounted above the resonator and heated the WGMR by illumination with several milliwatts of power. In this way, the resonator was locked to the free-running pump laser, which hence determined the system stability. By frequency-tuning the pump laser, we were thus able to change the temperature of the resonator and to fine-tune it to exactly degenerate operation, where both the pump and the alignment mode are simultaneously on resonance. The speed of the lock was limited by the intrinsic cooling rate of the resonator-coupler system.\\
In order to investigate the classical OPO behavior, we installed an electro-optic modulator for the pump beam, with which we swept the pump power faster than the locking cutoff frequency. We adjusted the modulator such that the pump power sweep spanned the threshold regime below and above.\\
The measurement port was used for coupling out parametric light and for determining the degenerate OPO threshold. The degenerate OPO threshold is lower than the thresholds of nondegenerate channels because the degenerate WGMs' overlap is the largest. Therefore, exciting the degenerate OPO above the threshold should normally prevent other channels from being similarly excited due to pump power clamping \cite{carlosnavarretebenllochthesis,PhysRevA.79.043820}. However, nondegenerate PDC channels may be excited when the pump is detuned from the WGM by several kilohertz \cite{Schunk:14} and can therefore be active due to the limited precision of the lock.\\
In order to distinguish the degenerate conversion channel from other excited channels, we swept the pump power, as explained before, and monitored an interference signal between the generated parametric light and the local oscillator in the detection unit. The local oscillator, which was derived from the fundamental of the pump laser, is naturally exactly degenerate. The interference signal was only observable when the frequency of the parametric light exactly matched the frequency of the local oscillator. The interference hereby tagged the degenerate conversion channel. For the lowest observed threshold signature, we fixed the pump detuning and the phase-matching detuning to zero and, by fitting Equation (\ref{eq:signalpower}) to the parametric signal, we determined the threshold value of the degenerate conversion channel.\\
To analyze the quadrature squeezing below the threshold, we used balanced homodyne detection implemented with two polarizing beam splitters and half-wave plates. The phase of the local oscillator was modulated with a piezo-mounted mirror. We monitored the interference signal with a direct difference photodetector on an electronic spectrum analyzer at an RF sideband frequency of 500 kHz with 100 kHz resolution bandwidth, 30 Hz video bandwidth, and a sweep time of 0.5 s. The detection efficiency of $\eta=86\%$ accrued from 97\% quantum efficiency of the photodiodes and 89\% beam path transmission.

\section{Experimental results}	

\subsection{Classical OPO behavior and threshold determination}
\begin{figure}[htbp]
\includegraphics{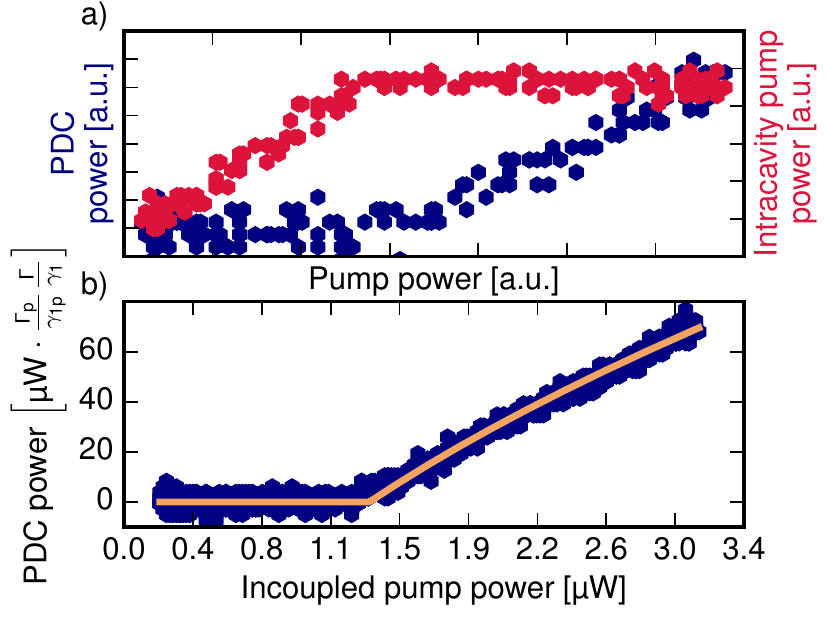}
\caption{a) Classical OPO behavior. Above the threshold, the intracavity pump power stays constant alongside increasing parametric power. b) Threshold of the OPO. Parametric light output in reflection of the first prism coupler versus pump power showing the square root behavior and the oscillation threshold described by Equation (\ref{eq:signalpower}). The fit yielded a threshold of 1.35 \textmu W.}
\label{fig:clamping}
\end{figure}

We extracted the classical OPO behavior by sweeping the pump power and detecting the parametric light after the first port and the intracavity pump power after the second port. As expected from the theory \cite{doi:10.1080/713820226}, we observed that the intracavity pump power first grows linearly with the incoupled pump power and then stays constant as soon as the OPO threshold is reached and bright parametric light is generated, which is shown in Figure \ref{fig:clamping}\, a).
\\We then determined the parametric oscillation threshold of the degenerate PDC channel, which depends on $\gamma_{1,2}$ according to Equation (\ref{eq:pumpthreshold}). The result of such a measurement using the experimental technique explained in the previous section is shown in Figure \ref{fig:clamping}\, b) and yielded a degenerate OPO threshold of 1.35 \textmu W.
This measurement technique worked up to the point when $2\Gamma\approx 2\pi\times 2.2\text{ MHz}$. For higher coupling rates, other nondegenerate PDC channels might have become more efficient, resulting in a lower respective threshold. Such a channel consequently causes pump clamping \cite{carlosnavarretebenllochthesis,PhysRevA.79.043820} by going above the threshold before the degenerate PDC channel. As a result, the degenerate channel can never reach its threshold. This affected our ability to perform the near-threshold measurements in a strongly overcoupled resonator (see Section 2), hence limiting the maximum squeezing we could observe for this experiment.\\
From several threshold measurements, we calculated the nonlinear coupling constant $g$ to be $2\pi\times(2.1\pm 0.2)\text{ kHz}$, which is an important figure of merit for quantum-optical applications of resonator-assisted PDC. Furthermore, we can use $g$ to calculate the expected OPO threshold for coupling conditions where the previously described measurement scheme breaks down.

\subsection{Squeezing generation}
\begin{figure}[htbp]
\includegraphics{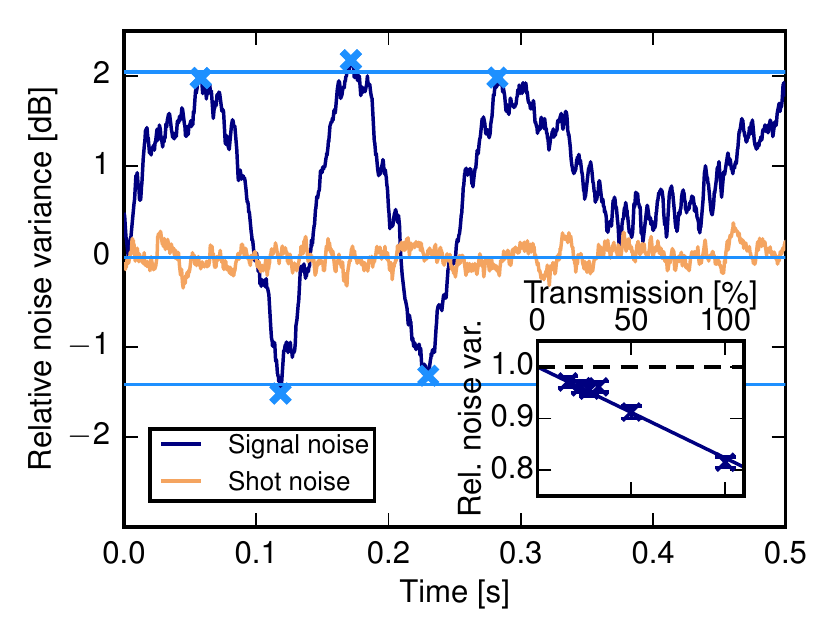}
\caption{Noise variance of the subharmonic field. Averaging the displayed extrema yields a squeezed noise variance of -1.4 dB and an anti-squeezed noise variance of 2.1 dB relative to shot-noise for 300 \textmu W of incident pump power at a sideband frequency of 500 kHz. The inset shows an attenuation measurement of a squeezed state verifying the linear regime of the detector.}
\label{fig:squeezing}
\end{figure}

In order to generate squeezing, the pump beam was frequency-locked to continuously excite the degenerate channel and the alignment beam was blocked. First, we calibrated the shot-noise reference level by measuring the quadrature noise of the local oscillator without interfering it with the signal field. Then, we introduced the signal field and detuned the pump, with linewidth precision, around the degenerate phase-matching point by changing the laser frequency and, consequently, the temperature of the locked resonator. In postprocessing, we selected the traces exhibiting the highest noise reduction, which corresponded to a setting of zero detuning from the degenerate phase-matching.\\
Far above the threshold, a bright signal field is generated that affects the shot-noise reference level, but this situation can be identified during postprocessing as the minima of the noise traces are shifted above the shot-noise level. Above but close to the threshold, the generated signal field is not bright enough to affect the reference level. Figure \ref{fig:squeezing}
shows a typical squeezing signature of such a measurement with a squeezed noise variance of $-1.4\pm 0.1$ dB and an anti-squeezed noise variance of $2.1\pm0.1$ dB for 300 \textmu W of incident pump power at an RF sideband frequency of 500 kHz. We estimated the OPO threshold with the given coupling rates and the coupling constant $g$ to be approximately 220 \textmu W with respect to 300 \textmu W of incident pump power. However, the expected signal power of about 1.5 \textmu W was three orders of magnitude smaller than the local oscillator and, consequently, negligible.\\
We verified the linear regime of the detector by attenuating the signal beam with neutral density filters and observing the linear decrease of both squeezing, which is shown in the inset of Figure \ref{fig:squeezing}, and anti-squeezing.\\
To test the theoretical model (\ref{eq:squeezingquadraturenoisevariance}), we needed to be able to measure and to reach the threshold power, and also to have control of the individual coupling rates $\gamma_1$ and $\gamma_2$. We conducted another measurement and changed the system such that it was operated in the slightly undercoupled regime with $2\Gamma=2\pi\times2.18\text{ MHz}$, where the parametric oscillation threshold was still measurable, and squeezing was measured at a sideband frequency of 2 MHz.
\begin{figure}[htbp]
\includegraphics{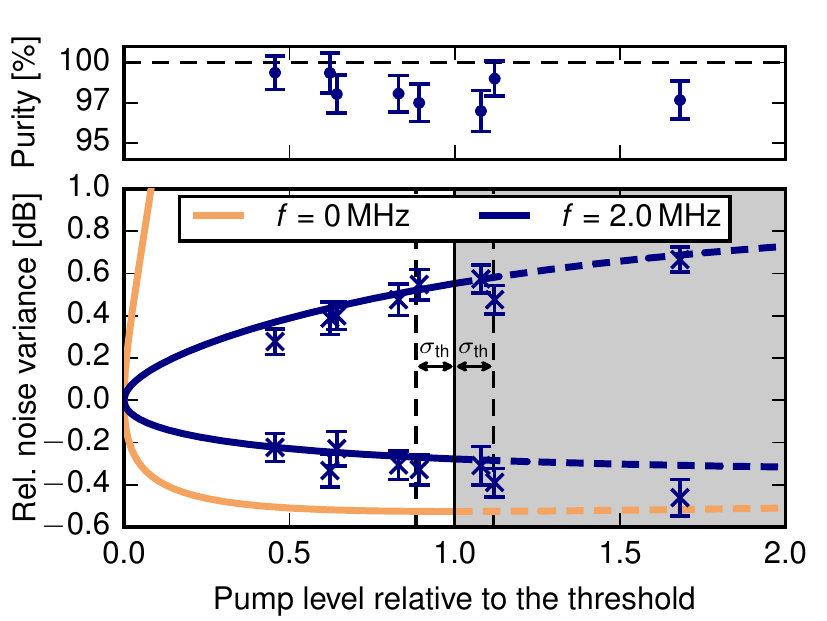}
\caption{Comparison of measured squeezing values with the theoretical model for below-threshold operation. The system was operated in the slightly undercoupled regime with $2\Gamma=2\pi\times2.18\text{ MHz}$. Though the theoretical model is incomplete close to and above the threshold, the model is consistent with the measurement.}
\label{fig:theoretical_model}
\end{figure}
For these parameters, the no-depletion model predicts the strongest squeezing at $\mathcal{G}\approx 4$, or at $P\approx 4P_\text{th}$, which is consistent with our result, shown in Figure \ref{fig:theoretical_model}. At the same time, we need to keep in mind that this model is inaccurate, as it disregards the pump clamping at the OPO threshold, clearly visible in Figure \ref{fig:clamping}\,a). To emphasize this limitation, we plot the theoretical curve above the threshold as a dashed line. It is interesting that even though incomplete, this model is still consistent with the above-threshold measurements in Figure \ref{fig:theoretical_model}. Below the measured threshold, the generated states were purely squeezed vacuum states without any bright signal power present. For the given measurement, we can theoretically expect a maximum amount of $-3$ dB of squeezing at threshold and at a sideband frequency of 2 MHz. For this estimate, we assumed a maximum coupling rate $2\gamma_2 \approx 2\pi \times 7.6\text{ MHz}$, which was separately verified as the maximum possible coupling rate for this particular experimental setup, and kept all other parameters of the theoretical curve of Figure \ref{fig:theoretical_model} such as $\mathcal{V}$ and $\gamma_1$, for example. As earlier mentioned in this section, we were limited by pump clamping and were consequently not able to experimentally access this coupling regime.\\
 Besides the quadrature variance, an important parameter characterizing squeezed states is the purity, defined for Gaussian states as $[\text{Var}(X_\text{sq})\text{Var}(X_\text{asq})]^{-0.5}$ \cite{PhysRevA.68.012314}. Equation (\ref{eq:squeezingquadraturenoisevariance}) implies that high purity can only be obtained when the second coupler dominates, i.e., $\gamma_2 \approx \Gamma$, which is also a condition required for generating maximum squeezing, as mentioned in the theoretical section. Purity is essentially the inverse of the area occupied by the quantum state in phase space. This area is minimal for pure states and is larger for mixed states. Therefore, achieving high purity is crucial in quantum communication and information applications of squeezed states, where decoherence and contamination with noise is highly detrimental. The squeezed vacuum states of light generated by our system featured a high purity, ranging from 97\% to 99\%. Perfect purity within the measurement error bars was demonstrated in a highly optimized monolithic cavity OPO \cite{PhysRevA.81.013814} for pump powers orders of magnitude higher than reported here but still far below the respective OPO threshold. We achieved comparably high purity below, at, and above threshold, as shown in Figure \ref{fig:theoretical_model}.

\section{Conclusion and outlook}
In this paper, we demonstrated the generation of squeezed vacuum states of light in a crystalline WGMR. We showed that this type of resonator can be used as a compact and power-efficient source, as we generated -1.4 dB  of squeezing at only 300 \textmu W of incident pump power. Our second-order nonlinear platform hereby offers a two orders of magnitude lower pump power than the recent third-order nonlinear on-chip approach \cite{arxiv:1904.07833}. Furthermore, we achieved genuinely single-mode vacuum squeezing in degenerate operation, which is available in the low megahertz regime. Compared to other highly engineered second-order nonlinear platforms, such as the monolithic PPKTP cavity approach \cite{PhysRevLett.104.251102,Yonezawa:10,Schonbeck:18}, our WGM-based approach does not produce a comparable amount of squeezing. However, we did not pursue competing with these highly optimized setups in terms of the mere amount of squeezing. Instead, we tried to build a compact and low-power squeezer that additionally has the known advantages of a crystalline WGMR, such as a tunable bandwidth within a regime compatible with atomic transitions and an extreme low pump power threshold.\\
We identified the technical constraints of our demonstration system that limited the observed squeezing and can mitigate them in future setups. The two key parameters are a high outcoupling rate $\gamma_2$ and a low incoupling rate $\gamma_1$, which can be considered as loss for squeezing through the first coupler. We can increase $\gamma_2$ by using a WGMR with a smaller major radius $R$ and larger $r$ \cite{Gorodetsky:99}, which facilitates higher overcoupling. The coupling at the first prism coupler can be drastically reduced by introducing polarization-selective coupling \cite{PhysRevApplied.7.024029}. Replacing the first diamond prism with a prism made of lithium niobate would prevent the outcoupling of the TM-polarized squeezed vacuum mode and consequently cancel $\gamma_1$. Such a technique has already been used together with a monolithic square resonator for generating squeezed vacuum states of light \cite{Brieussel:16}, but for pump powers two orders of magnitude higher. Finally, the competing nondegenerate modes can be eliminated by employing a single-mode WGMR, which is also a proven technology \cite{Savchenkov:06}.\\
We want to note that our system is a promising platform to study the squeezing process close to and above the threshold. Furthermore, we can scrutinize the interaction with adjacent PDC conversion channels leading to pump clamping.\\
In summary, this work illustrates that crystalline WGMRs constitute a compact, low-power source of squeezed vacuum states of light. Recent substantial progress in fabricating monolithic integrated lithium niobate microring resonators \cite{Zhang:17} makes the WGM-based platform a potential candidate for industrialized applications, as discussed in the introduction. Besides that, our system offers the possibility of combining the fields of optomechanics and nonlinear optics in just one monolithic device \cite{PhysRevLett.115.243603}, as high-Q crystalline microresonators have already been shown to provide optomechanical coupling \cite{PhysRevA.82.031804}.

\section*{Funding Information}
We acknowledge funding from the European Research Council (ERC) (PACART) and the Danish National Research Foundation, Center for Macroscopic Quantum States (bigQ, DNRF142). This work has been funded by the European Union's Horizon 2020 research and innovation program through the Quantum-Flagship project CiViQ (820466).

\section*{Acknowledgments}
The authors acknowledge the support from Lothar Meier and Oliver Bittel.

\bibliography{bibliography}

\end{document}